\documentstyle[12pt]{article}

\def\gsim{\raisebox{-4pt}{$\,\stackrel{\textstyle{>}}{\sim}\,$}}
\begin{document}
\begin{flushright}
\baselineskip=12pt
DOE/ER/40717--34\\
CTP-TAMU-50/96\\
ACT-15/96\\
\tt hep-ph/9610235
\end{flushright}

\begin{center}
\vglue 1.5cm
{\Large\bf No-scale supergravity confronts LEP diphoton~events} 
\vglue 1.5cm
{\Large Jorge L. Lopez$^1$, D.V. Nanopoulos$^{2,3}$, and A.~Zichichi$^4$}
\vglue 1cm
\begin{flushleft}
$^1$Department of Physics, Bonner Nuclear Lab, Rice University\\ 6100 Main
Street, Houston, TX 77005, USA\\
$^2$Center for Theoretical Physics, Department of Physics, Texas A\&M
University\\ College Station, TX 77843--4242, USA\\
$^3$Astroparticle Physics Group, Houston Advanced Research Center (HARC)\\
The Mitchell Campus, The Woodlands, TX 77381, USA\\
$^4$University and INFN--Bologna, Italy and CERN, 1211 Geneva 23, Switzerland\\
\end{flushleft}
\end{center}

\vglue 1cm
\begin{abstract}
We examine the possibility that some anomalous acoplanar diphoton events
observed at LEP may be consistent with the predictions of our previously proposed one-parameter no-scale supergravity model with a light gravitino, via the process $e^+e^-\to\chi\chi\to\gamma\gamma+{\rm E_{miss}}$. We find that
one such event may indeed be consistent with the model predictions for $m_\chi\approx(60-70)\,{\rm GeV}$. This region of parameter space is also consistent with the selectron and chargino interpretations of the CDF $ee\gamma\gamma+{\rm E_{T,miss}}$ event.
\end{abstract}

\vspace{1cm}
\begin{flushleft}
\baselineskip=12pt
October 1996\\
\end{flushleft}
\newpage
\setcounter{page}{1}
\pagestyle{plain}
\baselineskip=14pt

The puzzling $ee\gamma\gamma+{\rm E_{T,miss}}$ event observed by CDF \cite{Park} has sparked renewed interest in models of low-energy supersymmetry,
particularly those that predict photonic signals \cite{KaneDine,Gravitino,Events}. In this note we concentrate on a highly predictive one-parameter no-scale supergravity model \cite{One}\footnote{We
should clarify that the model considered in Refs.~\cite{One} has the same spectrum as the present model. However, the experimental signatures presented
in those papers differ from the present ones because the possibility of a light gravitino was not entertained at that time.} that realizes the light-gravitino scenario advocated previously as an explanation for the CDF event \cite{Gravitino}. The generic experimental signature for this model is ${\rm n}\gamma+{\rm E_{miss}}+X$. The number of observed photons may be n=0 in the case of gravitino pair production, n=1 in the case of gravitino-neutralino production, n=2 in the case of neutralino-neutralino production, and generally n=even (odd) if an even (odd) number of gravitinos and/or neutralinos are produced. The system $X$ may be `nothing' when only gravitinos and/or neutralinos are produced, but it may include visible particles when heavier supersymmetric particles are produced, as in the case of the CDF event, where $X=e^+e^-$ and n=2. 

In the context of LEP, the kinematically most accessible signals correspond to $X$=nothing and n=$1,2$, which result from gravitino-neutralino production
($e^+e^-\to\chi\widetilde G\to\gamma+{\rm E_{miss}}$) and neutralino-neutralino
production ($e^+e^-\to\chi\chi\to\gamma\gamma+{\rm E_{miss}}$). As of the end
of LEP161 running, there are no reported excesses in the single-photon cross
section. There exist however interesting acoplanar diphoton events, at least
in the OPAL data sample taken at $\sqrt{s}=130\,{\rm GeV}$ (${\cal L}\approx3\,{\rm pb}^{-1}$) \cite{OPAL} and the L3 data sample at $\sqrt{s}=161\,{\rm GeV}$ (${\cal L}\approx11\,{\rm pb}^{-1}$) \cite{L3}.
In view of this situation, one may presume that single-photon events are not being observed, while diphoton events may be starting to show up. This scenario
is in fact one of two mutually exclusive ones (the other being the converse),
as we have recently demonstrated in the context of general light-gravitino models \cite{1vs2gamma}. This dichotomy is possible, in spite of the ample kinematical accessibility of the single-photon process ($m_\chi<\sqrt{s}$),
because the single-photon cross section can be suppressed below the 0.1~pb level by choosing a suitable gravitino mass ($m_{\widetilde G}>3\times10^{-5}\,{\rm eV}$), while the diphoton cross section remains unaffected. Further support for this scenario may come from LEP runs at higher center-of-mass energies.

\begin{figure}[p]
\vspace{6in}
\includegraphics{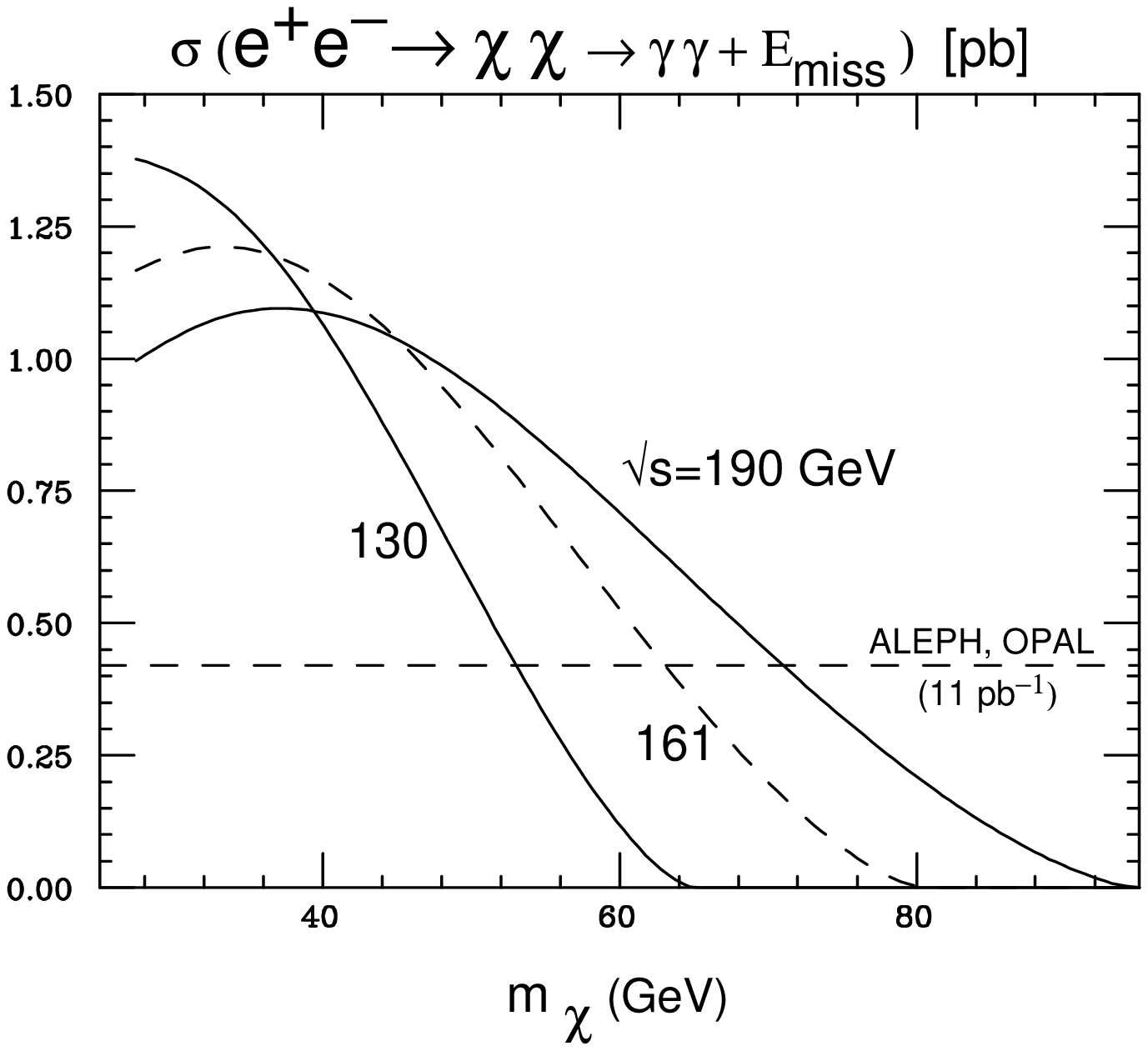}
\caption{Diphoton cross section versus neutralino mass at various LEP center-of-mass energies in no-scale supergravity with a light gravitino. The preliminary ALEPH and OPAL upper limits obtained at LEP161 are indicated.}
\label{fig:N1N1-updated}
\end{figure}

Let us start by displaying in Fig.~\ref{fig:N1N1-updated} the diphoton cross section as a function of the neutralino mass for various LEP center-of-mass energies, where we have made use of the result $B(\chi\to\gamma\widetilde G)\approx1$ \cite{Events}. It is important to realize that, besides the neutralino mass ($m_\chi$), this cross section also depends on the neutralino composition and the selectron masses ($m_{\tilde e_{L,R}}$). These additional variables are not independent parameters in the model, but instead vary continously with $m_\chi$: $m_{\tilde e_L}\sim 1.5\, m_{\tilde e_R}\sim2 m_\chi$, while the neutralino approaches a pure bino composition at high neutralino masses. (The dependence of the diphoton cross section for generic values of the parameters is studied in Ref.~\cite{1vs2gamma}.) There have been several upper limits imposed on the diphoton cross section by the various LEP Collaborations at LEP~1.5 \cite{OPAL,DELPHI} and LEP161 energies \cite{DELPHI,ALEPH,Oct8}. Of these the preliminary ALEPH and OPAL LEP161 upper limits of $0.42\,{\rm pb}$ (obtained with ${\cal L}\approx 11\,{\rm pb}^{-1}$ \cite{Oct8}) are most constraining in our model. Fig.~\ref{fig:N1N1-updated} shows that $m_\chi\gsim60\,{\rm GeV}$ is required.  Corresponding diphoton limits from LEP~1 are not available, and in any event would imply $m_\chi>{1\over2}M_Z$ at best, a sensitivity already surpassed at LEP161.

We now turn to the analysis of the three acoplanar diphoton events that have
been observed to have characteristics not fit well by the expected
$e^+e^-\to\nu\bar\nu\gamma\gamma$ background. Most notably, the missing invariant mass $M_{\rm miss}=\sqrt{E^2_{\rm miss}-p^2_{\rm miss}}$
(with $E_{\rm miss}=\sqrt{s}-E_1-E_2$ and $p_{\rm miss}=|\vec p_1+\vec p_2|$) is peaked around $M_Z$ for the background,
whereas it has no particular structure for the diphoton signal \cite{KaneDine}. The kinematical information on the three events is listed in Table~\ref{Table1}, from where we can see that $M_{\rm miss}$ is indeed substantially removed from the $Z$ peak in all cases. We should remark that we are fully aware that these events may well be due to the expected background \cite{Pohl}. However, we find it well motivated to seek alternative explanations for these events. This is particularly relevant in the context of
the model of Refs.~\cite{Gravitino,Events} where the diphoton cross section is at the same level as the expected background. We should add that both OPAL and L3 have additional acoplanar diphoton events with $M_{\rm miss}\approx M_Z$, which have been disregarded in the present analysis because of their very likely background origin.

\begin{table}[t]
\caption{The kinematical information on the three acoplanar photon pair events
observed by OPAL at LEP~1.5 and L3 at LEP161. Also shown is the missing invariant mass $M_{\rm miss}$. All momenta, masses, and energies in GeV, angles in radians.}
\label{Table1}
\begin{center}
\begin{tabular}{lccrrrccc}
&$\sqrt{s}$&$E_1$&$E_2$&$\cos\theta_1$&$\cos\theta_2$ &$\phi_1$&$\phi_2$
&$M_{\rm miss}$\\ \hline
OPAL&$130.26$&$28.5$&$18.4$&$0.473$&$-0.926$&$4.06$&$1.17$&$81.3$\\
L3a&$161.00$&$12.9$&$37.9$&$0.548$&$0.872$&$5.41$&$3.32$&$101.4$\\
L3b&$161.00$&$36.2$&$19.3$&$0.690$&$-0.836$&$1.99$&$4.43$&$102.9$\\
\hline
\end{tabular}
\end{center}
\hrule
\end{table}

Assuming that these diphoton events are the result of the underlying
$e^+e^-\to\chi\chi\to\gamma\gamma\widetilde G\widetilde G$ process, one can
perform a Monte Carlo simulation of the unseen momenta (that of the two
gravitinos) and obtain distributions of neutralino masses that are consistent
with the kinematics of the events. Essentially one varies the 3 components
of one of the gravitino 3-momenta and obtains the neutralino mass. Momentum
conservation determines the other gravitino 3-momentum and energy conservation
determines its energy. Demanding an essentially massless gravitino gives
a constraint, as does the requirement that the second neutralino reconstruct
to the same mass as the first neutralino. Altogether one is left with a single
free parameter. This procedure has been described in detail in Ref.~\cite{Events}. One finds that energy and momentum conservation in the events restrict the range of possible neutralino masses to:
\begin{eqnarray}
{\rm OPAL:\quad}&16.6\,{\rm GeV}<m_\chi<52.5\,{\rm GeV}
\label{eq:OPAL}\\
{\rm L3a:\quad}&39.8\,{\rm GeV}<m_\chi<58.2\,{\rm GeV}
\label{eq:L3a}\\
{\rm L3b:\quad}&13.7\,{\rm GeV}<m_\chi<46.3\,{\rm GeV}
\label{eq:L3b}
\end{eqnarray}
Given the above lower limit on the neutralino mass from direct searches 
($m_\chi\gsim60\,{\rm GeV}$) it appears unlikely that the OPAL
and L3b events may be of neutralino origin.

\begin{figure}[p]
\vspace{6in}
\includegraphics{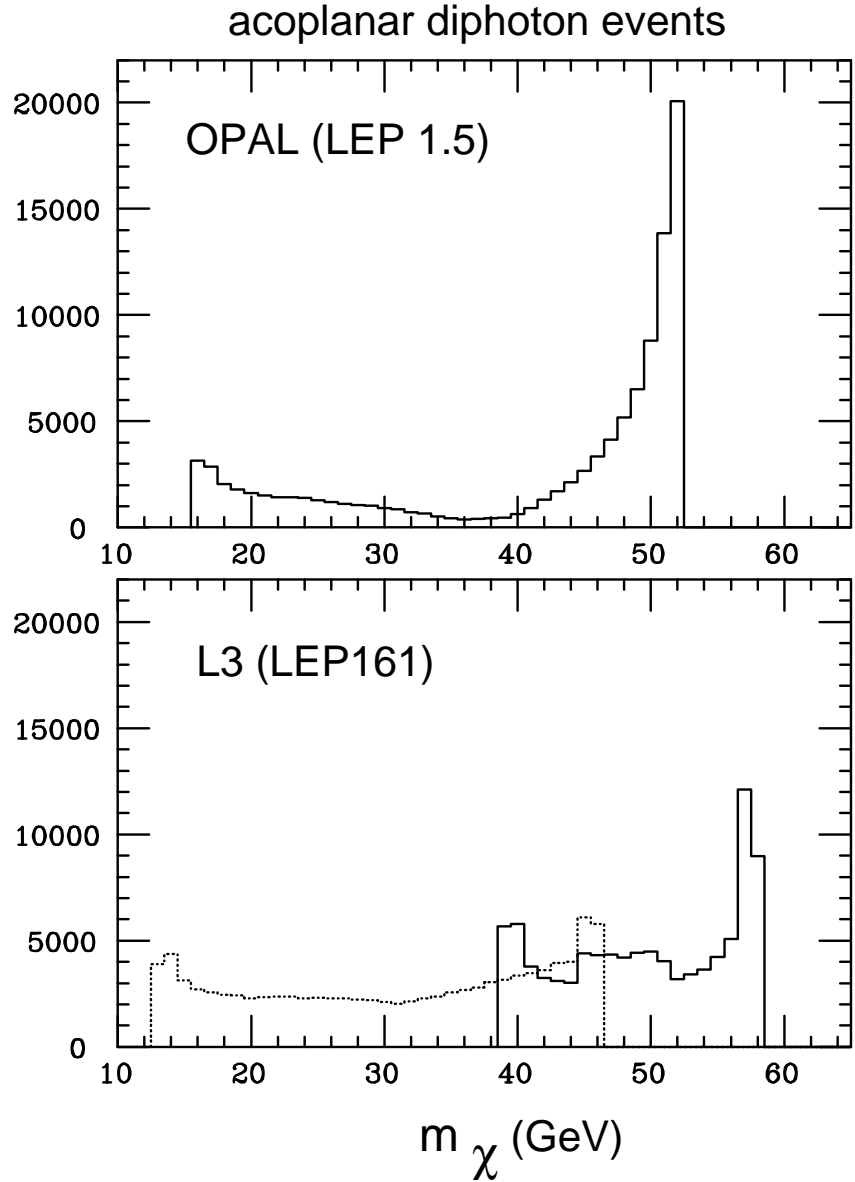}
\caption{Neutralino mass distributions consistent with the kinematics of
three anomalous acoplanar diphoton events observed at LEP, under the assumption of $e^+e^-\to\chi\chi\to\gamma\gamma\widetilde G\widetilde G$ as the underlying
process.}
\label{fig:bins-updated}
\end{figure}

More information, although of a statistical nature, can be obtained by examining the distributions of neutralino masses that are obtained. These are shown in Fig.~\ref{fig:bins-updated}. (The distribution upper and lower limits agree with those in Eqs.~(\ref{eq:OPAL},\ref{eq:L3a},\ref{eq:L3b}).) Let us first note the distinct preference for neutralino masses near the upper limit of the allowed mass range. The figure makes it also apparent that a common source for these events is not inconceivable, and in fact quite natural at around $m_\chi\approx50\,{\rm GeV}$. However, as noted above, the lower limit on the neutralino mass seemingly forces us to disregard the OPAL and L3b events. We should point out that the neutralino mass range allowed in the L3a event is subject to some uncertainty that will broaden the distribution somewhat. For instance, the neutralino masses need not be exactly the same on account of the finite neutralino width or because near threshold one neutralino may be produced off-shell. Also, the experimental momenta used in our calculations have some uncertainties that are not available and have not been accounted for in our simulation. To be conservative one may speculate that neutralino masses above 60 GeV, and as high as 70 GeV may be consistent with the data. Clearly more data are needed to reach a more definitive conclusion.

Let us now consider the rates for these events. The OPAL event was observed at $\sqrt{s}=130\,{\rm GeV}$ in ${\cal L}\approx2.8\,{\rm pb}^{-1}$ of data.
The cross section for $m_\chi=50\,(60)\,{\rm GeV}$ is $\sigma^{\rm 130}=0.55\,(0.10)\,{\rm pb}$, entailing $1.5\,(0.3)$ expected events --
not inconsistent with observations except for the direct mass limit. The L3 events were observed at $\sqrt{s}=161\,{\rm GeV}$ in ${\cal L}\approx11\,{\rm pb}^{-1}$ of data. The cross section for $m_\chi=50,60,70\,{\rm GeV}$ is $\sigma^{\rm 161}=0.88,0.51,0.18\,{\rm pb}$, entailing $9.7,5.6,2.0$ expected events respectively -- seemingly preferring the upper end of the allowed neutralino mass range. It is interesting to note that the signal cross section
is at the same level as the expected Standard Model background.

\begin{figure}[p]
\vspace{6in}
\includegraphics{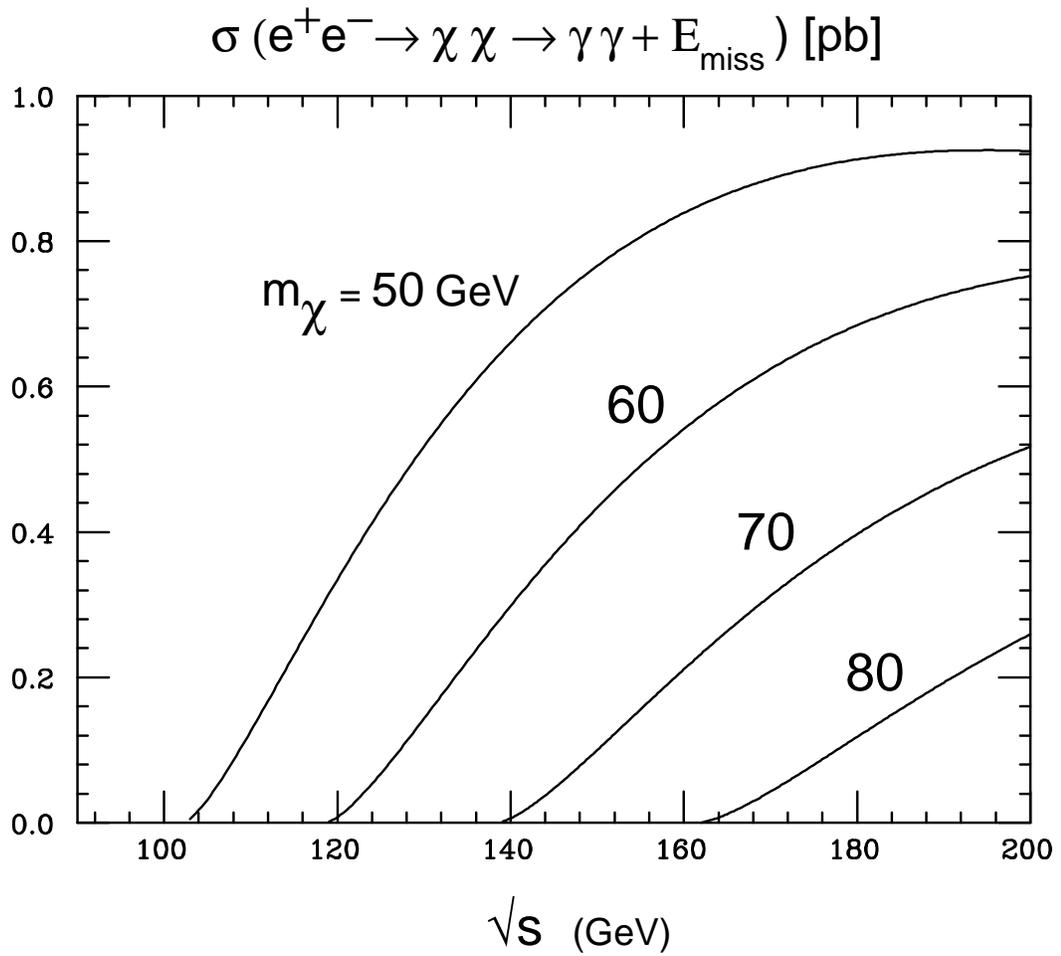}
\caption{Diphoton cross section versus LEP center-of-mass energy for various
neutralino masses in no-scale supergravity with a light gravitino.}
\label{fig:VaryRoots}
\end{figure}

\begin{figure}[p]
\vspace{6in}
\includegraphics{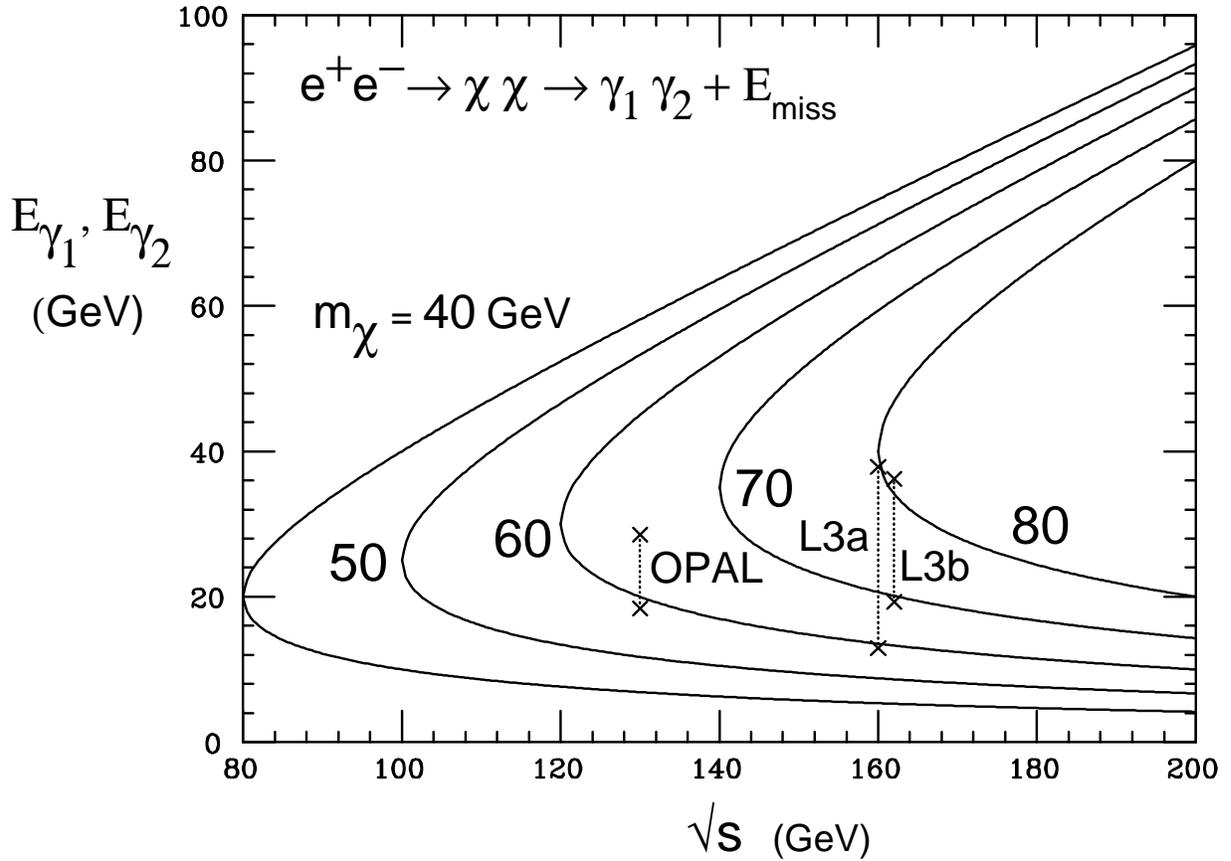}
\caption{Range of photonic energies as a function of center-of-mass energy for fixed values of the neutralino mass, as would be expected from neutralino pair-production events at LEP. The OPAL and L3 events are indicated. Upper
limits on $m_\chi$ values consistent with the kinematics of the events can
be deduced from this figure.}
\label{fig:gammas-updated}
\end{figure}

For future reference, we have also calculated the diphoton cross section as
a function of the center-of-mass energy, for selected neutralino masses, as
shown in Fig.~\ref{fig:VaryRoots}. For instance, for $m_\chi=60\,{\rm GeV}$,
the cross section increases from 0.51~pb to 0.61~pb when going from $\sqrt{s}=161\,{\rm GeV}$ up to $\sqrt{s}=175\,{\rm GeV}$, a 20\% increase.
Kinematically speaking, the photon energies in the diphoton events are
restricted to the range ${\sqrt{s}\over4}(1-\beta)<E_{\gamma_1,\gamma_2}<
{\sqrt{s}\over4}(1+\beta)$, with $\beta=\sqrt{1-4m^2_\chi/s}$. We have found
it useful to display the kinematical information of the diphoton events in the backdrop of these ranges of photonic energies as a function of $\sqrt{s}$, for fixed values of $m_\chi$. This plot is shown in Fig.~\ref{fig:gammas-updated}, where the OPAL and L3 events are indicated. From the plot one can read off the
maximum neutralino masses that are consistent with the kinematics, in agreement
with Eqs.~(\ref{eq:OPAL},\ref{eq:L3a}). The range in the L3b event is more
restrictive than this plot would imply (see Eq.~(\ref{eq:L3b})) because momentum conservation is quite restrictive in this case, whereas Fig.~\ref{fig:gammas-updated} only includes energy conservation.
Hopefully future interesting acoplanar diphoton events will fall in this plot forming a coherent pattern. We should emphasize that the results in Figs.~\ref{fig:bins-updated} and \ref{fig:gammas-updated} are model-independent consequences of the kinematics of the events, whereas Figs.~\ref{fig:N1N1-updated} and \ref{fig:VaryRoots} are dynamical predictions valid only in our present model.

Next we consider the overlap between the region of parameter space
apparently preferred by the LEP diphoton events ({\em i.e.}, the L3a event) with that consistent with an explanation for the CDF $ee\gamma\gamma+{\rm E_{T,miss}}$ event. This event may be explained as selectron pair production ($p\bar p\to\tilde e^+\tilde e^-\to e^+e^-\chi\chi\to e^+e^-\gamma\gamma\widetilde G\widetilde G$), in which case $m_\chi>68\,{\rm GeV}$ appears required \cite{Events}. Alternatively, the event may be explained as chargino pair-production  ($p\bar p\to\chi^+_1\chi^-_1\to
(e^+\nu_e\chi)(e^-\bar\nu_e\chi)\to e^+e^-\nu_e\bar\nu_e\gamma\gamma\widetilde G\widetilde G$), in which case $m_\chi>55\,{\rm GeV}$ appears required \cite{Events}. Thus, the parameter space preferred by the diphoton event,
when relaxed somewhat to account for the possible uncertainties ({\em i.e.},
$m_\chi\approx(60-70)\,{\rm GeV}$) appears consistent with both interpretations of the CDF event. The region of parameter space apparently singled out by LEP corresponds to $m_{\tilde e_R}\approx(95-105)\,{\rm GeV}$ and
$m_{\chi^\pm_1}\approx(110-130)\,{\rm GeV}$. The selectron cross section at the Tevatron in the selected selectron mass range is $(6$--$8)\,{\rm fb}$ \cite{Events}, consistent with observations. In the chargino interpretation there are many additional processes that one has to contend with \cite{Events},
and a detailed experimental analysis of the detection efficiencies needs to
be performed. We can ascertain though that the total diphoton plus $\rm E_{T,miss}$ signal from such sources is $(1$--$2)\,{\rm pb}$. This
signal however receives a significant contribution from the process $3\ell+\gamma\gamma+{\rm E_{T,miss}}$, which is not likely to have a high detection efficiency, as earlier studies involving traditional supersymmetric trilepton searches have shown \cite{trileptons}. The first searches for such inclusive signal at CDF have recently been reported \cite{Toback}, although no explicit limits have yet been released.

We look forward to runs at LEP~2 with increased statistics and/or larger center-of-mass energies, as they would have to yield many acoplanar
diphoton events should this model describe Nature.

\newpage
\section*{Acknowledgements}
The work of J.~L. has been supported in part by DOE grant DE-FG05-93-ER-40717. The work of D.V.N. has been supported in part by DOE grant DE-FG05-91-ER-40633.

\clearpage

\clearpage

\clearpage

\clearpage
\end{document}